\documentclass[showpacs,preprintnumbers,amssymb]{revtex4}
\usepackage{graphicx}% Include figure files
\usepackage{dcolumn}% Align table columns on decimal point
\usepackage{bm}% bold math
\def\be{\begin{equation}} 
\def\ee{\end{equation}}
\def\bea{\begin{eqnarray}} 
\def\eea{\end{eqnarray}}
\def\line{\hbox to \hsize}    
\def\frac #1#2{{#1\over #2}}

\def\tr{{\rm  tr\,}}

\def\psid{\psi^{\dagger}}

\def\hPsi{\hat \Psi}
\def\Psid{\Psi^{\dagger}} 
\def\hPsid{\hat \Psi^{\dagger}}

\def\hH{{\hat H}}

\def\Det{{\rm Det\,}}

\def \x{{\bf x}}
\def \y{{\bf y}} 
\def \z{{\bf z}}     

\def \d{{\bf d}}
\def \k{{\bf k}}

\def\vev #1{{\langle #1\rangle}}
\def\1{\mbox{\bf 1}}

\begin{document}
%\draft %(only for revtex) 

\title{Edge modes, edge currents, and gauge invariance 
in $p_x+ip_y$ superfluids and superconductors.
}

\author{ MICHAEL STONE}

\affiliation{University of Illinois, Department of Physics\\ 1110 W. Green St.\\
Urbana, IL 61801 USA\\E-mail: m-stone5@uiuc.edu}   

\author{RAHUL ROY}

\affiliation{University of Illinois, Department of Physics\\ 1110 W. Green St.\\
Urbana, IL 61801 USA\\E-mail: rahulroy@uiuc.edu}

\begin{abstract}

The excitation spectrum of a  two-dimensional  $p_x+ip_y$
fermionic superfluid, such as a thin film of $^3$He-A,
includes  a gapless Majorana-Weyl fermion which is confined
to the boundary   by Andreev reflection. There is also a
persistent ground-state boundary current which  provides a
droplet containing  $N$ particles with angular momentum 
$\hbar N/2$.  Both of these boundary effects are associated
with bulk Chern-Simons effective actions. We show that the 
gapless edge-mode is required for the gauge invariance of
the total effective action, but the same is not true of the
boundary current.

\end{abstract}

\maketitle

%\pacs{PACS numbers:    }
%\newpage 

\section{Introduction}

Read and Green \cite{read-green} have pointed out close 
parallels between two-dimensional \hbox{$p_x+ip_y$} chiral
fermionic superfluids (a thin film of  $^3$He-A for
example) and the $\nu=1/2$ Pfaffian quantum Hall state. 
The many-body ground-state wavefunction of both systems
contains   a  Pfaffian factor, and both support gapless
Majorana-Weyl edge modes \cite{volovik97}. In addition, 
both systems can host vortex defects with non-Abelian
statistics \cite{moore-read,ivanov01}.   The   $p_x+ip_y$
superfluid  also possesses an equilibrium edge current that
is consistent with an  $\hbar$-per-Cooper-pair intrinsic
angular momentum \cite{ishikawa77,mermin-muzikar,kita98}.
This current is similar to that induced by a confining
potential at the boundary of a Hall droplet, and occurs
because chiral fermionic superfluids manifest an analogue
of the quantum Hall effect, even in the absence of an
external magnetic field \cite{volovik88}.

The principal  differences between the Pfaffian Hall state
and the chiral superfluid are that, in addition to  the
Pfaffian, the  Hall-state wavefunction includes a symmetric
$\nu =1/2$ Laughlin factor; also the Hall system is spin
polarized, while the superfluid retains two active spin
components. A less significant  difference  is that the
Hall fluid is incompressible while a  neutral superfluid
like $^3$He-A  has a gapless acoustic mode. This acoustic 
mode will, however, be gapped in a charged superfluid
state, such as that believed to exist in the layered
Sr$_2$RuO$_4$ superconductors  \cite{rice95,baskaran96}.

In the quantum Hall effect, the lack of low energy bulk
degrees of freedom leads to the dynamics of the boundary 
and the bulk being  intimately connected. Currents flowing
in the  the bulk serve to soak up the anomalies of the
edge-mode   conservation laws \cite{stone91}, and, more
profoundly, the bulk many-body wavefunction can be
constructed from the conformal blocks of the edge conformal
field theory  \cite{moore-read}.  The existence of the
conformally invariant Majorana edge mode suggests that
something similar is true for the superfluid. 

Because of the  interest of  quantum systems with
non-Abelian statistics, both for their intrinsic appeal as
exotic physics and for their potential application to
quantum computing \cite{kitaev97,ioffe02}, there is reason
to search  for the simplest possible description of their
dynamics. Thus we seek an  effective action
which captures all the essential low-energy degrees of
freedom of the system.  The construction of such an  
effective action for the fermionic Pfaffian Hall state is,
however, rather complicated \cite{fradkin-nayak-tsvelik-wilkzek}. It  is
therefore worth  examining  other systems with similar properties. 

One such system is the {\it bosonic\/} Pfaffian quantum
Hall state, which can perhaps be realized in rotating Bose
condensates. The bosonic Pfaffian state has a ground-state
wavefunction consisting of the product of the
antisymmetric  Pfaffian and an antisymmetric $\nu=1$
Laughlin factor, making it even under  particle
interchange.  In this 
case \cite{fradkin-nayak-tsvelik-wilkzek,fradkin-nayak-schoutens}
the  effective action is an $SU(2)$ Chern-Simons term  at
level $k=2$. This means that  the edge states  form
representations of an  ${SU(2)}_2$  current  algebra.
Indeed, the   space of low energy  states is spanned by
wavefunctions consisting of polynomials in the complex
coordinates $z_i$ that  vanish when any three  of the $z_i$
coincide.  The generating  function  for  the number of
such polynomials is a  character of the ${SU(2)}_2$
algebra \cite{stoyanovskii}.

The $p_x+ip_y$ superfluid is also relatively easy to
analyze because  its effective action has been  computed  by
standard gradient expansion methods
 \cite{volovik88,volovik-yakovenko89}. It is the purpose of
this paper to explore some  of the mathematical and
physical aspects  of the  resulting expression. We
will see  that  the  response of the fluid to a non-Abelian   gauge field
which couples  to the spin degree of freedom is governed by
a Chern-Simons term of conventional form, but with a
coefficient corresponding to level 
$k={\textstyle \frac12}$. Since invariance under
``large'' gauge transformations demands that the level $k$ be an
integer, this means that the
spin action cannot be  gauge invariant on its own. Gauge
invariance is restored only when we take into account the
role of  the spin-triplet order parameter in spontaneously
breaking of the $SU(2)$ symmetry  down to $U(1)$, and then
by the subsequent absorption of the residual $U(1)$ gauge
dependence by the  $U(1)$ axial anomaly of   the chiral
edge-states. The topological origin  of the  edge modes is
therefore illuminated.   The response to   an
Abelian particle-number gauge field is governed  
by an action  which looks superficially
Chern-Simons like,  but is deficient in that it lacks the
terms with time derivatives. In this case, gauge
invariance   is  immediately manifest  once we include the
$U(1)$  Goldstone mode in the effective action
 \cite{volovik88}. In particular, no boundary  degree of
freedom is required. Although this means that there is no
explicit coupling of the Abelian gauge field to the edge
modes,  the Abelian part of the effective action does
describe the  Hall-like   edge {\it current\/}, which
therefore has a different origin from the edge {\it modes\/}. It
also reveals an essential difference between    the real  Hall
effect and its field-free analogue.  In the Hall effect the
current is proportional to the applied  electric field. In
the superfluid,  the absence of the $\dot {\bf A}$ in the
action  means  that  the induced current is proportional 
not to the electric field, but to  the  gradient of the
fluid density. The ``Hall current'' should therefore  be
thought of as a two-dimensional analogue of  the
Mermin-Muzikar current \cite{mermin-muzikar}, which is due
to the intrinsic angular momentum of the fluid. 

There have been several recent papers discussing the
effective actions for chiral superfluids and 
superconductors \cite{goryo-ishikawa,furusaki}, but our point
of view differs from  these in its interpretation of the
``Hall effect", the magnitude and origin of the
edge-current, and in its emphasis on the role of gauge
invariance in establishing the bulk/edge connexion.    

In the next section we will discuss the Bogoliubov
action for the planar $p_x+ip_y$ spin-triplet superfluid.
In section  \ref{SEC:boundary_current} we will find the
eigenfunctions of the corresponding Bogoliubov-de Gennes
Hamiltonian in  rigid walled containers, and use  them to
compute the edge-mode spectrum and the magnitude of the
persistent edge-currents. In section four we will describe
the effective actions governing the response of the fluid
to external gauge fields  that couple to particle number
and to spin. We then  show how the existence of the
persistent  boundary current  and gapless edge-modes can be
deduced from the bulk effective action.

\section{Bogoliubov-de Gennes operator}

The fermionic part of the action describing a BCS superconductor
can be written in Nambu two-component formalism as     
\be
S= \int d^2xdt\left(\frac 12 \Psid(i\partial_t -
\hH)\Psi\right).
\label{EQ:bogoliubov-action}
\ee 
Here,
\be
\Psi =\left[\matrix{\psi_\alpha\cr \psid_\alpha}\right],\quad
\Psi^\dagger =\left[\matrix{ \psid_\alpha, & \psi_\alpha}\right],
\ee
are Grassmann fields with a spinor index
$\alpha=\uparrow,\downarrow$, and 
$\hH$ 
is the Bogoliubov-de Gennes Hamiltonian
\be
\hH= \left[\matrix{ \hat h & \hat  \Delta \cr
           \hat  \Delta^\dagger & -\hat h^T}\right].
\label{EQ:bogoliubov-hamiltonian}
\ee
The entries in $\hH$ are single-particle operators acting on the 
tensor product of the position and spinor spaces. 
The hermiticity of $\hH$ requires that the 
single-particle Hamiltonian $\hat h$ be Hermitian, and Fermi
statistics requires that the gap function $\hat \Delta$  be 
skew-symmetric, in this combined space. 

For a
Galilean invariant fluid of particles of mass $m$, the
Hamiltonian is
\be
\hat h= \left(-\frac 1{2m} \nabla^2 -\epsilon_f\right)I,
\ee
where $I$ is the identity operator in spin space and
$\epsilon_f$ the Fermi energy. For a
superconductor $\hat h$ can be a  more general function,
$\epsilon(\hat {\bf p})I$, of $\hat {\bf p}=-i\nabla$. 

The gap function will usually be a dynamical field, and
the  total action will contain additional terms that serve
to determine its value {\it via\/} a gap equation. We are,
however, interested only in the response of the fermions to
prescribed changes in the gap function $\hat \Delta$, so we
will not need to make these terms explicit. Further, we  are
primarily interested in topological effects, so the precise
form of  $\hat \Delta$ is  not significant as long as 
it produces the required symmetry breaking.
For a $p_x+ip_y$, spin-triplet superfluid, such as
$^3$He-A, we may take 
\be 
\hat \Delta = \frac 12
\left(\frac{\Delta}{k_f}\right)e^{i\Phi/2} \left\{\hat
\Sigma, \hat P\right\}e^{i\Phi/2}. 
\ee Here, 
$\{\phantom\sigma,\phantom\tau\}$ denotes an
anticommutator, $\Delta$ is the magnitude of the induced gap
in the quasiparticle spectrum, and $\Phi$ is  the overall phase
of  the order parameter.  The spin part $\hat \Sigma$  is a
symmetric $2\times 2$  matrix   with entries  
\be
\Sigma_{\alpha\beta}= (i (\sigma\cdot
\d)\sigma_2)_{\alpha\beta}, 
\ee 
where $\d$ is a unit
vector. The orbital part of the gap function   is contained
in the operator  $\hat P$, which we will take  to be   
\be
\hat P=-i(\hat p_x+i\hat p_y)\equiv
-(\partial_x+i\partial_y), 
\ee  
corresponding to  Cooper
pairs with their $l=1$  angular momentum vector $\bf l$
directed in the $+\hat \z$ direction, {\it
i.e.}~perpendicular to  the plane of the fluid.  This  
orientation for ${\bf l}$ ensures that the entire
Fermi-surface is gapped. We will fix   ${\bf l}=\hat \z$
throughout  this paper.  We will also take the magnitude of
the gap $\Delta$ to be a constant. Although this parameter
should  be determined self-consistently through a  gap
equation, its magnitude serves only to provide an upper
limit for what  we mean by ``low-energy" degrees of
freedom, and so any  variation has no role in the following
discussion.

When $\hat \Sigma$ and $\Phi$ are
constants, the operator ordering of the spin, phase, and
orbital parts  of $\hat \Delta$ is unimportant. When these
quantities  vary in space, however, we need the  
anticommutator of $\hat \Sigma$ and $\hat P$, and the
symmetric distribution of the overall phase $e^{i\Phi}$
about it, to ensure the  antisymmetry of $\hat \Delta$. 

In the sequel, our calculations will be strictly 2+1
dimensional. They therefore apply to a single stratum of a
layered superconductor, or, for  a thin film of $^3$He-A,
to the case when only a single transverse momentum mode
lies below the Fermi surface. When $n$ transverse modes are
occupied but the order-parameter remains independent of
$z$, our effective actions  should be multiplied by $n$.

\section{Edge modes and boundary currents}
\label{SEC:boundary_current}

We now suppose the superfluid to be confined by  rigid
walls at which the wavefunction is required to vanish. 
We also assume, for the duration of this section,   that the
spin vector $\d$  lies in the $\hat \y$ direction, 
so that $\hat \Sigma=iI$ and the two spin components decouple.

\subsection{Rectangular Geometry}
\label{SEC:planar}

If we substitute 
\be
\Psi=\left[\matrix{a\cr b}\right]e^{ik_x x +ik_y y}, 
\ee 
with constant $a$, $b$,
into the Bogoliubov equation, $\hH\Psi=E\Psi$, the
eigenvalue condition  reduces to  
\be
\left[\matrix{\epsilon(k) & \left(\frac
k{k_f}\right)e^{i\theta}\Delta \cr
                 \left(\frac k{k_f}\right)e^{-i\theta}\Delta &
		 -\epsilon(k)}\right]
 \left[\matrix{a \cr
               b }\right] 
 = E \left[\matrix{a \cr
                   b }\right].
\ee
Here  $\theta$ is the polar angle such that  $k_x=k\cos\theta$,
$k_y=k\sin\theta$. The plane-wave eigenstates are therefore 
\be
\Psi_{E,\k} = e^{i\sigma_3 \theta/2}
\frac{1}{2\sqrt{E(E+\Delta)}} 
\left[\matrix{E+\epsilon(k)+\Delta\cr
E-\epsilon(k)+\Delta}\right]e^{ik_x x +ik_yy}
\ee
with   
$E=+\sqrt{\epsilon^2(k)+\Delta^2(k^2/k_f^2)}$, and
\be
\Psi_{-|E|,\k} = e^{i\sigma_3 \theta/2}
\frac{1}{2\sqrt{|E|(|E|-\Delta)}} 
\left[\matrix{|E|-\Delta-\epsilon(k)\cr
|E|-\Delta+\epsilon(k)}\right]e^{ik_x x +ik_yy}
\ee
with   $E=-\sqrt{\epsilon^2(k)+\Delta^2(k^2/k_f^2)}$.
We will usually be interested in states
close to the Fermi surface where we can approximate 
the energy as  
\be
E=\pm\sqrt{v_f^2(k-k_f)^2+\Delta^2},
\ee
where $v_f$ is the Fermi-velocity.
We have 
considered only one spin component. There are
really  two sets of such solutions, one for spin up and one for
spin down.

If we expand the quantized field operator $\hat\Psi$ in terms of the plane wave
states
\be
\hat \Psi =\left[\matrix{\hat \psi\cr \hat
\psi^\dagger}\right]=\sum_{E,\k} \hat a_{E,\k}\Psi_{E,\k}
\ee
then, in order for the upper and lower components to
reconstruct $\hat\psi$ and $\hat \psid$ respectively, the
operators $\hat a_{E,\k}$ must obey 
the reality condition $\hat a_{E,\k} = \hat
a_{-E,-\k}^\dagger$.

\subsubsection{Edge states}

\begin{figure}
\includegraphics[width=2.5in]{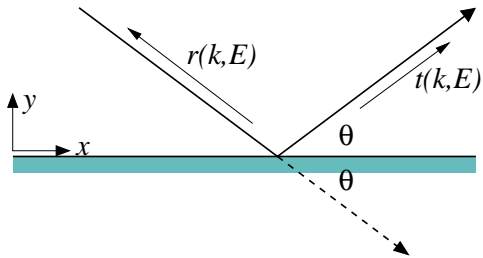}
\caption{\label{wall}Geometry of specular and Andreev
 reflection from the boundary.}
\end{figure}

%\vbox{
%\vskip 20 pt
%\centerline{\epsfxsize=2.5in\epsffile{wall.eps}}
%{\narrower\smallskip\centerline{\sl Geometry of specular and Andreev
%reflection from the boundary. }
%}}

We now investigate the effect of a boundary.
Suppose, as shown in figure \ref{wall}, that there is a  wall at $y=0$, with the fluid lying above
it. We  seek solutions to the Bogoliubov equation in the
form
\be
\Psi=C_+\left[\matrix{a_+(y)\cr b_+(y)}\right]e^{ik_f x \cos\theta  
+ik_f y\sin\theta } 
-C_-\left[\matrix{a_-(y)\cr b_-(y)}\right]e^{ik_f x\cos\theta  
-ik_f y\sin\theta },
\ee
where now $a_\pm$ and $b_\pm$ are allowed to  vary with $y$ --- but
slowly  on the scale of 
$k_f$.
We impose the boundary condition that $\Psi=0$ at $y=0$.
The resultant equations for $a_\pm$ and $b_\pm$ coincide
with those derived in the appendix for the  solutions of a 
one-dimensional Dirac hamiltonian. The vanishing of the
wavefunction at the wall becomes (because of the minus sign
before $C_-$) the continuity of the Dirac wavefunction at
$x=0$. The  incoming particle therefore sees the reflecting
boundary only as an abrupt  change  in the phase of the
orbital part of the order parameter, which  jumps   from
$-\theta $ to $+\theta$.    The  transmitted Dirac wave 
with amplitude $t(k,E)$ corresponds to those particles that
have been  specularly  reflected at  boundary, and the
backscattered  Dirac wave with amplitude $r(k,E)$
corresponds to those particles that have been  Andreev
reflected, and so retrace their path.  The only
significant  differences  between the two dimensional
geometry  and the one-dimensional problem solved in the
appendix    are that  the co-ordinate ``$x$'' in the
appendix  is the distance along the classical trajectory,
and that there we have set the Fermi velocity $v_f$ to
unity. Thus, expressions appearing in the appendix, such
as  $e^{-\kappa |x|}$ must here  be  replaced by
$e^{-\kappa |y|/(v_f\sin(\theta))}$. 

On 
making the appropriate translations from the results in the
appendix, we find that we have a  bound state
\be
\chi_0(k_x)= \sqrt{\frac{\Delta}{2v_f}}e^{ik_f x\cos\theta } 
\sin (k_f y\sin\theta )
e^{-\Delta y/v_f}\left[\matrix{1\cr
1}\right]
\ee
with energy
\be
E_0(k_x)= -\Delta\cos\theta= -\Delta(k_x/k_f).
\label{EQ:edge_spectrum}
\ee
 The
associated edge excitations  are therefore {\it chiral\/}, or {\it
Weyl\/}, fermions, having group velocity $\partial E/\partial
k_x$ only in the  $-\hat\x$  direction. This motion is in the
opposite sense to the  anticlockwise orbital
angular momentum of the Cooper pairs. 

These edge modes have a topological character. All that is
required for them to exist is
that the fluid density fall to zero at the
boundary \cite{volovik97}. Indeed the edge-mode
spectrum (\ref{EQ:edge_spectrum}) for our rigid wall
boundary  coincides with that  obtained by Read and
Green \cite{read-green} in the  opposite extreme of a very
soft wall.  

The contribution of the edge-modes to the field operator
$\hat\Psi$ is
\be
\sum_{k_x} \hat a_{0,k_x}\chi_{0,k_x}(x),
\ee
and, in  order for the upper and lower components to
reconstruct $\hat\psi$ and $\hat \psid$ respectively, we
must have $\hat a_{0,k_x} = \hat
a_{0,-k_x}^\dagger$. The edge-mode field
\be
\hat\Psi_0(x) = \int_{-k_f}^{k_f}\frac{dk}{2\pi} \hat
a_{0,k_x}\chi_{0,k_x}(x) 
\ee
is therefore {\it real\/}, $\hat\Psi_0(x)= \hat\Psi_0^\dagger(x)$, or 
{\it Majorana\/}.

It is difficult for  anything interact with a  Majorana-Weyl
particle. A Weyl
particle can interact only {\it via\/}  currents, and  no
current operator can be constructed
from a single Majorana field.
In our superfluid there  are, however, {\it two\/}  Majorana-Weyl fields: one for spin
up and one for spin down. Out of these two real
fields we can construct one complex Weyl field
\bea
\hPsi_c&=& \frac 1{\sqrt 2}(\hPsi_{\uparrow 0}+i\hPsi_{\downarrow
0}),\nonumber\\
\hPsid_c&=& \frac 1{\sqrt 2}(\hPsi_{\uparrow 0}-i\hPsi_{\downarrow
0}),
\eea
and from this
we can construct a unique $U(1)$ current operator
\be
\hPsid_c\hPsi_c= \frac {i}{2}\left(\hPsi_{\uparrow 0}\hPsi_{\downarrow 0}-
\hPsi_{\downarrow 0}\hPsi_{\uparrow 0}\right) 
=\frac 12 \hPsi_0 \sigma_2\hPsi_0.
\ee
The edge modes may therefore  interact with  the $\sigma_2$
component of a spin-coupled gauge field.

\subsubsection{Boundary current}

The doubling of the number of degrees of freedom in  the
Bogoliubov formalism requires us,  when computing
ground-state expectation values of an operator,  to  sum
the contributions of all occupied (negative energy) 
states, but then divide by two to prevent over-counting --- 
this division by two being equivalent to imposing the
reality condition on the $\hat a_{E,{\bf k}}$.  The mass
current carried by the state  
\be 
\Psi=\left[\matrix{a\cr
b}\right]e^{ik_fx \cos\theta  +ik_f y\sin\theta }, 
\ee
where $a$, $b$ are slowly varying compared to $k_f$ is
therefore  \be {\bf j}=\frac 12(|a|^2+|b|^2)(k_f
\cos\theta, k_f\sin\theta). \ee 

The edge modes  with $k_x>0$ have negative energy, and  so
are occupied in the ground state where they   make a
positive contribution  to the boundary momentum density.
For a Galilean invariant system, the momentum density is
also the mass current.  The edge modes therefore tend to
produce a  boundary current that flows  in   the same
sense  as the Cooper-pair  rotation. The bound states,
however,  are not the only contribution to this boundary
current. The balance is   provided  by the phase-shifted 
scattering states, which (as they do the theory of
fractional charge \cite{goldstone-wilczek,stone85}) partially
cancel  the contribution  of a bound state when it has
negative energy and is occupied, and partially make up for 
the absence of its contribution  when it has positive
energy and is unoccupied. 

The current in the  direction perpendicular to the wall will cancel between the
incoming and outgoing waves\footnote{We ignore the cross terms which
fluctuate as $\exp i2k_f$, and contribute no net flux},  
but the $\hat \x$ components will add.
Using the results from the appendix, we find that 
the net mass-current  running near the boundary is (for a
single spin component) 
\bea
\int _0^\infty j_x dy  &=&\int_0^1 \frac{d(k_f\cos\theta)}{2\pi}
\left(\frac{\theta}{2\pi}\right) k_f\cos \theta +\int_{-1}^0 
\frac{d(k_f\cos\theta)}{2\pi}
\left(\frac{\theta}{2\pi}-\frac 12 \right) k_f\cos \theta
\nonumber\\
&=& 2\int_0^1 \frac{d(k_f\cos\theta)}{2\pi}
\frac{\theta}{2\pi} k_f\cos \theta\nonumber\\
&=& \frac{k_f^2}{16\pi} =\frac{1}{4}\rho,
\eea 
where $\rho=k_f^2/4\pi$ is the number density per spin component.
If this current  flows  at the edge of a disc-shaped
region of radius $R$, it provides angular momentum
\be
{\bf L}= 2\pi R^2\left(\frac{ \rho}{4}\right) \hat \z =\frac N 2
\hat \z.
\ee
This result agrees with that of Kita \cite{kita98}, and
Volovik \cite{volovik92} and  is exactly the angular momentum  we would 
expect of a fluid of tightly bound Cooper pairs, each pair
having orbital  angular momentum of $\hbar\hat \z$. 
That the same
result holds in the weak coupling limit, where only the
particles near the Fermi-surface are affected by the
pairing, is more surprising, and  it 
is often referred to as the ``angular momentum
paradox'' (see Kita \cite{kita98} for a recent review of this).  

Our result for the boundary current differs from that of
Furusaki {\it et al.\/} \cite{furusaki}, because they consider
only the the bound state contributions.
The boundary current arising from the bound
states alone is (again for  a single spin component)
\be\frac 12 \int_0^1
\frac{d(k_f\cos\theta)}{2\pi}
 k_f\cos \theta= \frac{k_f^2}{8\pi} =\frac{\epsilon_f m}{4\pi}.
\ee
After multiplying by two to take into account the two spin components,
this
coincides with equation (2.8) in Furusaki {\it et
al.\/} \cite{furusaki}, 
and differs from  the actual persistent boundary 
current   by a factor of two.

We have computed the boundary current only in the weak
coupling limit, but the result that it precisely accounts
for  the $\hbar$-per-pair angular momentum should remain
valid even as the coupling is increased. This is because,
provided the ground state evolves adiabatically, its total
angular momentum  cannot change as we alter parameters in
the Hamiltonian. Adiabatic evolution may fail if there is
spectral flow, and such flow does occur and change the
angular momentum 
when we compute the angular momentum of some vortex
configurations in bulk $^3$He-A  \cite{volovik95}, and also when we consider
the {\it dynamical\/} intrinsic angular momentum
density \cite{volovik-mineev}, but in the present case
spectral  flow can only occur through the gapless edge
states and the $E_0(k_x)=-E_0(-k_x)$ Majorana symmetry
prevents the spectrum moving {\it en-masse\/} with respect
to the chemical potential. For the same reason, 
the total boundary current should also
not be affected by local deviations of the order parameter 
away from its bulk form. The ``fractional charge''
interpretation \cite{goldstone-wilczek,stone85,stone-gaitan}
of the current ensures this, because the
total fractional charge depends only on the asymptotics of the
order parameter, and is unaffected by local variations.  

\subsection{Disc geometry}

When the fluid has disc geometry it is convenient to 
use polar coordinates $r$, $\theta$, and to work with
angular momentum eigenstates $\propto e^{il\theta}$. For
example, we have     
\be
-\nabla^2 J_l(kr)e^{il\theta}= k^2 J_l(kr)e^{il\theta}, 
\ee  
where $J_l(kr)$ is a Bessel function.
In polar coordinates, the orbital part of the
gap operator  $\hat P=-(\partial_x+i\partial_y) $ becomes    
\be
\hat P=-e^{i\theta}\left(\frac {\partial}{\partial r} +\frac i r\frac 
{\partial}{\partial \theta}\right),
\ee   
and has a particularly simple action on the Laplace eigenfunctions:
\be
\hat P e^{il\theta}J_l(kr) = k e^{i(l+1)\theta}J_{l+1}(kr).
\ee
The adjoint operator  
\be
\hat P^\dagger = -\left(-\frac 1 r\frac { \partial }{\partial
r}r +\frac i r\frac 
{\partial}{\partial \theta}\right)e^{-i\theta}= -
e^{i\theta}\left(-\frac {\partial}{\partial r} +\frac i r\frac 
{\partial}{\partial \theta}\right),
\ee 
similarly acts to reduce the angular momentum eigenvalue
\be
\hat P^\dagger e^{il\theta}J_l(kr) = k e^{i(l-1)\theta}J_{l-1}(kr).
\ee

We look for 
eigenstates in the form
\be
\psi= \left[\matrix{a J_{l+1}(kr)e^{i(l+1)\theta}\cr
                    b  J_{l}(kr)e^{il\theta}}\right].
\ee
The Bogoliubov equation, $\hH\Psi=E\Psi$, then reduces to 
\be
\left[\matrix{\epsilon(k) & (k/k_f)\Delta \cr
                 (k/k_f)\Delta&
		 -\epsilon(k)}\right]
 \left[\matrix{a \cr
               b }\right] 
 = E \left[\matrix{a \cr
                   b }\right].
\ee
 The  eigenstates are
therefore  
\be
\Psi_{E,l}(r,\theta)= 
\frac{1}{2\sqrt{E(E+\Delta)}}
		  \left[\matrix{(E+\epsilon+\Delta) 
		  e^{i(l+1)\theta}J_{l+1}(kr)\cr
                  (E-\epsilon
		 +\Delta)e^{il\theta}J_{l}(kr)
		 }\right], 
\ee
where the energy eigenvalue  is  
\be
E=+ \sqrt{\epsilon^2(k)+ (k/k_f)^2\Delta^2},
\ee
and 
\be
\Psi_{-|E|,l}(r,\theta)
= \frac{1}{2\sqrt{|E|(|E|-\Delta)}}
\left[\matrix{(|E|-\Delta-\epsilon)e^{i(l+1)\theta}J_{l+1}(kr) \cr
                (|E|-\Delta+\epsilon|E|)e^{il\theta}J_{l}(kr)
		 }\right], 
\ee
with energy 
\be
E=-\sqrt{\epsilon^2(k)+ (k/k_f)^2\Delta^2}.
\ee 

As before, we are  interested in momenta near the
Fermi-surface where the energy  can be approximated by
\be
E= \pm \sqrt{v_f^2 (k-k_f)^2+ \Delta^2}.
\ee

\begin{figure}
\includegraphics[width=2.20in]{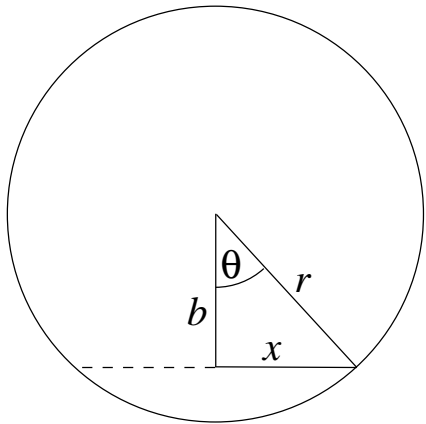}
\caption{\label{WKB} The geometry of the WKB
approximation to the Bessel function.}
\end{figure}

When we confine the fluid by imposing  rigid wall boundary conditions at $r=R$, the
eigenstates will be  a linear combination
\be
\Psi_l=C_+\left[\matrix{a_+J_{l+1}(k_+r)e^{i(l+1)\theta}\cr
                     b_+ J_{l}(k_+r)e^{il\theta}}\right]
-C_-\left[\matrix{a_-J_{l+1}(k_-r)e^{i(l+1)\theta}\cr
                     b_- J_{l}(k_-r)e^{il\theta}}\right],
\ee
of unconfined states  with slightly different momentum
$k_\pm=k_f\pm k$,
but a common energy  
\be
E=\pm \sqrt{v_f^2k^2 +\Delta^2}.
\ee 
The coefficients $a_\pm$ and $b_\pm$ are given by
\be
\left[\matrix{a_\pm\cr
              b_+}\right]= \left[\matrix{E\pm v_fk+\Delta\cr
             E\mp v_fk+\Delta }\right].
\ee
To examine  the consequences of the  condition that $\Psi=0$ at
$r=R$ we use the WKB approximation for the
Bessel function
\be
J_l(kr) \approx \sqrt{\frac{2}{\pi
kx(r)}}\sin(kx(r)-l\theta(r)
-\pi/4), \qquad r \gg b.
\ee
Here $x(r)$ and $\theta(r)$ are functions of $r$, defined in
terms of the semi-classical impact parameter,  $b= l/k$,
by $x=r\sin\theta$ and $b=r\cos\theta$. As illustrated
 in figure \ref{WKB}, the parameter   $x$ has the physical
interpretation of being the distance along the
straight-line semi-classical trajectory.  This approximation
is  quite accurate once   $r$ exceeds $b$ by more than a
few percent, and  is therefore reliable except for a few 
large  values of $l$ which correspond to classical 
trajectories grazing the boundary.  Using the WKB
approximation and the explicit form of the coefficients
$a_\pm$ and $b_\pm$ we we end up with  exactly the same
equations for the bound state and $S$ matrix that we found
in  the planar boundary case. 

%\vbox{
%\vskip 20 pt
%\centerline{\epsfxsize=2.20in\epsffile{WKB.eps}}
%{\narrower\smallskip\centerline{\sl The geometry of the WKB
%approximation to the Bessel function.} }
%}

There is some 
advantage to  working with a circular container, however.
With a finite length boundary, the set of edge modes becomes
discrete ---  being labeled by the integer $l$.
The  angle $\theta$  between the semi-classical trajectory and the boundary
is  $\theta(R)$, and  $l= k_fR\cos\theta(R)$.
In terms of $l$, therefore, the bound state has energy
\be
E_0(l) \approx -\left(\frac{l}{l_{{max}}}\right)\Delta,
\ee
where $l_{{max}}=k_fR$. 
The WKB approximation is not quite  good quite good enough to 
distinguish between $l$ and $l+\frac 12$ in this expression, 
but on general grounds we know that if $(u,v)^T$ is an
eigenstate of the Bogoliubov Hamiltonian  with energy $E$, then
$(v^*,u^*)^T$ is also an eigenstate with energy $-E$. Under
this transformation $\Psi_l\to \Psi_{-(l+1)}$, and so the
correct equation must be
\be
E_0(l) = -\left(\frac{l+1/2}{l_{{max}}}\right)\Delta.      		     
\ee
There is therefore no exact zero-mode. 
If, however, the bulk fluid were to contain an odd number of vortices,
and thus  the phase of the order parameter  wind an
odd number of times as we encircle the boundary, then the
$l$'s appearing in the upper and lower components of
$\Psi$  would  differ by an {\it even\/} number. In this case
there  {\it will\/} be  a zero energy edge state. Since each vortex
has a  zero-mode in its core \cite{ivanov01},  there will be an unpaired  zero-energy core 
mode which can  pair with the zero-energy edge state
and  so preserve the even dimension of
the total Bogoliubov-particle Hilbert space.

This dependence of the edge-mode spectrum on  the number of
the bulk vortex excitations is reminiscent  of what happens 
in the quantum Hall effect. There the Hilbert space sector
of the edge conformal field theory also depends on the number and
type of vortex quasi-particles that are present in the bulk. 

\section{Effective actions}

We now discuss the effective action governing the low-energy
dynamics of the superfluid. We will obtain the action by
examining the response of the fluid to external gauge fields
that couple to the particle number and spin of the fluid.

\subsection{Particle-number symmetry}

We   begin by gauging the $U(1)$ symmetry corresponding to
particle-number conservation. We will again hold the $\d$
vector fixed in  the $\hat \y$ direction so we can treat
each spin  component separately. We  minimally couple the
particle-number current to  an Abelian gauge
field $(A_0,{\bf A})$, where $A_0$ is the time  component
and ${\bf A}\equiv(A_1, A_2)$ are the in-plane components of
the
externally imposed field. This requires   the  replacement   
\be
i\partial_t- \hH \to i\partial_t -\hH(A,\Phi)
\ee
where
\be 
\hH(A,\Phi)=\left[\matrix{
 -\frac1{2m}(\nabla -i{\bf A})^2-A_0, & 
 i\left(\frac{\Delta}{k_f}\right)e^{i\Phi/2}
\hat Pe^{i\Phi/2}\cr
 -i\left(\frac{\Delta}{k_f}\right)e^{-i\Phi/2}
\hat P^\dagger e^{-i\Phi/2}, &
\frac1{2m}(\nabla
+i{\bf A})^2+A_0}\right].
\ee
The resulting  action 
\be
S(A,\Phi,\Psi,\Psid)=\int d^2x\,dt\,  \Psid\left(i\partial_t
-\hH(A,\Phi)\right)\Psi
\ee   
is then invariant under the local gauge
transformation
\be
\left[\matrix{\psi\cr \psid}\right]\to 
\left[\matrix{e^{i\phi}\psi,\cr
e^{-i\phi}\psid}\right],
\label{EQ:Abelian_fermi_gauge}
\ee
provided we simultaneously  transform
\bea
\Phi&\to&  \Phi+2\phi,\nonumber\\
{\bf A} &\to& {\bf A}+ \nabla\phi,\nonumber\\
 A_0 &\to& A_0 +\partial_t \phi.
\label{EQ:Abelian_bose_gauge}
\eea
Because the Grassmann measure in the path integral is left invariant
by the transformation (\ref{EQ:Abelian_fermi_gauge}), 
the effective action
\bea
iS_{\rm eff}^{\rm num}(A,\Phi)&=& 
\ln\left\{\int
d[\Psi]d[\Psid]\exp[iS(A,\Phi,\Psi,\Psid)]\right\}
\nonumber\\
&=&\frac 12 \ln\Det(i\partial_t-
\hH(A,\Phi)) 
\label{EQ:Abelian_determinant}
\eea 
will be invariant under the transformation
(\ref{EQ:Abelian_bose_gauge}).
If we compute  (\ref{EQ:Abelian_determinant}) to
second order in the gauge field and its gradients,
we find \cite{volovik88,goryo-ishikawa}, for a single  spin
component, 
\bea
S_{\rm eff}^{\rm num}(A,\Phi)&=& 
\int d^2xdt\left\{ \frac {\rho_0}{2m} \left[\frac 1{c_{s}^2} 
\left(\frac{\partial \Phi/2}{\partial t}
-A_0\right)^2-
\left(\nabla \Phi/2-{\bf A}\right)^2 \right] \right.\nonumber\\
&&\quad \left.-\sigma_{xy}\left(\frac{\partial \Phi/2}{\partial t}
-A_0\right)\left(\nabla\times {\bf A}\right)_z - \rho_0\left(\frac{\partial \Phi/2}{\partial t}
-A_0\right)\right\}.\nonumber\\  
\label{EQ:Abelian_action}
\eea
Here,   $\sigma_{xy}=\frac
1{8\pi}$, the parameter $c_s$ is the speed of sound, and $\rho_0$ the
equilibrium number density.  The coefficient $\sigma_{xy}$
is topologically stable, but the other quantities depend on
the details of the fluid \cite{volovik88}.
For a 2+1 dimensional Galilean invariant system of particles with mass $m$ we have $\rho_0=
\frac{m}{2\pi} \epsilon_f$ and  $c_s=v_f/2$. The
action (\ref{EQ:Abelian_action})  is manifestly
invariant under the transformation 
(\ref{EQ:Abelian_bose_gauge}). 

The term with coefficient $\sigma_{xy}$ contains  a
Chern-Simons-like part 
\be
\sigma_{xy}\int d^2x\,dt  \epsilon_{0ij}A_0\,\partial_iA_j. 
\ee
This  is not a complete Chern-Simons action, however, because
there is no $\epsilon_{i0j}A_i\,\partial_tA_j$ term. It
does,
nonetheless, imply the existence  of a Hall-like response to the
external field. We find for the particle-number current 
\bea
{\bf j}_{\,\rm num}&\equiv & \frac{\delta S_{\rm
eff}^{\rm num}}{\delta \bf A}=
\frac{\rho_0}{m} 
\left(\nabla \Phi/2-e{\bf A})\right) +\sigma_{xy} 
({\hat
\z}\times \nabla)\left(\frac{\partial \Phi/2}{\partial t}
-A_0\right)\nonumber\\
&=& \rho_0 {\bf v}_s +\sigma_{xy} ({\hat
\z}\times \nabla)\left(\frac{\partial \Phi/2}{\partial t}
-A_0\right).
\eea
Although the term with $\sigma_{xy}$ contains a gradient
of $A_0$, it is  not equal to $\sigma_{xy}
({{\bf E}\times \hat\z})$, where ${\bf E}=\nabla A_0-\dot{\bf
A}$, as it would be in the Hall effect.
(Observe that $\nabla\dot \Phi/2$ cannot be 
$\dot {\bf A}$ in disguise, because  the former is necessarily curl-free.) 
We note, however, that the combination
$(\partial_t \Phi/2
-A_0)$ occurs in the the expression for the density 
\be
\rho \equiv  \frac{\delta S_{\rm eff}^{\rm num}}{\delta A_0}= \rho_0 - \frac
{\rho_0}{mc_s^2}\left(\frac{\partial \Phi/2}{\partial t}
-A_0\right)+ \sigma_{xy} \nabla\times {\bf A}.
\ee
Consequently, it seems preferable to write  
\be
{\bf j}_{\,\rm num}= \rho_0{\bf v}_s- \frac 1{4m} 
({\hat
\z}\times \nabla)(\rho -\sigma_{xy} B_z),
\ee
where $B_z=(\nabla\times{\bf A})_z$, and so recognize that 
the ``Hall''  current depends on the external field 
primarily through its effect in modifying the density. The
natural analogy is then with the  bound current ${\bf
j}_{\,\rm bound}
= \nabla\times {\bf M}$ in a magnet with varying
magnetization ${\bf M}$. 
In the superfluid, the   magnetic-moment density ${\bf
M}$ is replaced by the  intrinsic angular momentum
density $\frac 12 \hbar(\rho-\sigma_{xy} B_z)\hat \z$. The
$\sigma_{xy} B_z$ correction is presumably present 
because a  diamagnetic response to the external  field will 
reduce   the {\it kinetic\/} angular momentum of a Cooper
pair even while leaving its {\it canonical\/} angular
momentum fixed at $\hbar$. In the absence of an external
$B_z$ field, we therefore have a mass flux 
\be
{\bf j} = m\,{\bf j}_{\,\rm num}= \frac 14 \nabla \times {\rho\hat \z},
\ee
and this is   a planar analogue of the
Mermin-Muzikar current \cite{mermin-muzikar}. When  $\rho$
goes to zero slowly at a boundary we can use this
Mermin-Muzikar expression to compute the equilibrium 
boundary-current. The resulting   boundary momentum
density  coincides with that we computed for  a rigid wall
in  section \ref{SEC:boundary_current}, and again provides
an $\hbar$-per-Cooper pair angular momentum. This agreement
between the rigid-boundary calculation and the gradient
expansion when expressed in terms of the density, coupled
with  the physical explanation in terms of the intrinsic
angular momentum of the fluid, leads us to conjecture  
that the total mass-current is always proportional to  the
change in density, rather than to the  external force that 
causes the  change.

\subsection{Spin-rotation symmetry}

The  fields in    
\be
S= \int d^2xdt\left(\Psid(i\partial_t - \hH)\Psi\right)
\ee
can also be  coupled to an $SU(2)$ gauge field which acts on the
spin indices. To do this we replace the derivatives in $S$ 
by covariant derivatives  
\be
\partial_\mu \to \partial_\mu +{\mathcal A}_\mu,
\ee
where
\be
{\mathcal A}_\mu = i\sigma_a {\mathcal A}^a_\mu
\ee
is an externally imposed gauge field.
Under a local gauge  transformation the Fermi fields transform as
\be
\Psi= \left[\matrix{\psi\cr \psid}\right]\to 
\left[\matrix{U\psi\cr U^*\psid}\right],\quad 
\Psid=\left[\matrix{\psid,&\psi}\right]
\to
 \left[\matrix{\psid U^{-1},&\psi U^T}\right],
\label{eq:su2_gauge_transform}
\ee
where $U\in SU(2)$. The covariant derivatives transform
as   
\be
\partial_\mu +{\mathcal A}_\mu\to U^{-1}(\partial_\mu
+{\mathcal A}_\mu)U=
\partial_\mu + (U^{-1}{\mathcal A}_\mu U+U^{-1}\partial_\mu U).
\ee

When considering
how the  
transformation act in  the $-\hat h^T$ entry in $\hH$, 
we need to recognize  that   derivative operators appearing there 
are  the transpose  $\partial_\mu^T=-\partial_\mu$ of
those in $\hat h$, and so the
covariant derivatives will be also be the transpose 
 $\partial_\mu^T+{\mathcal A}_\mu^T =
-\partial_\mu+{\mathcal A}_\mu^T$.
Using 
$U^*=(U^{-1})^T=(U^T)^{-1}$, we have  
\bea
U^T(\partial_\mu^T +{\mathcal A}_\mu^T)(U^T)^{-1}&=&
-\partial_\mu + U^T{\mathcal A}_\mu^T(U^T)^{-1}- U^T(U^T)^{-1}\partial_\mu U^T
(U^T)^{-1}\nonumber\\
&=&  -\partial_\mu + (U^{-1}{\mathcal A}_\mu U+ U^{-1} \partial_\mu
U)^T\nonumber\\
&=&[\partial_\mu + (U^{-1}{\mathcal A}_\mu U+ U^{-1} \partial_\mu U)]^T,
\eea
and the the effect of the transformation is consistent for  both
entries:
\be
{\mathcal A}_\mu\to A_\mu^U \equiv U^{-1} {\mathcal A}_\mu U +U^{-1}\partial_\mu U. 
\ee
Note that $({\mathcal A}^U)^V={\mathcal A}^{UV}$. 
For the off diagonal terms we have  
\be
U^{-1}\hat \Sigma U^* = U^{-1}\left(i(\sigma\cdot {\bf
d})\sigma_2\right)U^*
= \left(iU^{-1}(\sigma\cdot {\bf d})U\sigma_2\right).
\ee
The net result is that the gauged action is invariant under
the transformation (\ref{eq:su2_gauge_transform}), provided
we simultaneously transform
\bea
({\bf d}\cdot\sigma)&\to&  U(\sigma\cdot {\bf
d})U^{-1}\nonumber\\
{\mathcal A}_\mu &\to& {\mathcal A}_\mu^{U^{-1}} =
U{\mathcal A}_\mu U^{-1} + U\partial_\mu U^{-1}.
\eea

Following Volovik and Yakovenko
 \cite{volovik-yakovenko89}, we now compute the effective action
with the  $\bf d$ vector fixed to lie in the $\hat \y$
direction, and  find
a Chern-Simons term
\be
S_{\rm eff}^{\rm spin}({\bf d}=\hat \y, {\mathcal A})=\frac 1{8\pi} \int
\tr\left({\mathcal A}d{\mathcal A}+\frac 23
{\mathcal A}^3\right).
\ee
Here we use differential-form notation  in which ${\mathcal
A}\equiv  i\sigma_a
{\mathcal A}^a_\mu  dx^\mu$ is a matrix  valued 1-form.
For any compact simple  gauge-group $G$, the  Chern-Simons action
is defined to be
\be
C({\mathcal A})= \frac 1{4\pi} \int_\Omega \tr\left({\mathcal
A}d{\mathcal A}+\frac 23
{\mathcal A}^3\right),
\ee
where  ${\mathcal A}\equiv  i\lambda_a {\mathcal A}^a_\mu  dx^\mu$ is
a  ${\rm
Lie\,}(G)$-algebra-valued
1-form, and, as is customary, we 
normalize the trace and the Hermitian generators
$\lambda_a$ by $\tr(\lambda_a\lambda_b)=2\delta_{ab}$.
When such a term is to appear  in a functional integral, 
\be
Z= \int d[{\mathcal A}]e^{ikC({\mathcal A})},
\ee
then
coefficient $k$ must  be quantized. This is because under a
gauge transformation
\be
{\mathcal A}\to{\mathcal  A}^g \equiv g^{-1}{\mathcal A}g+g^{-1}dg
\ee
we have
\be
C({\mathcal A})\to C({\mathcal A}^g)=C({\mathcal A}) -\frac{1}{12\pi} \int_\Omega\tr 
\left[(g^{-1}dg)^3\right] -\frac{1}{4\pi}\int_{\partial
\Omega}\tr(dg g^{-1}{\mathcal A}).
\label{eq:polyakov}
\ee 
Here, $\Omega$ is the region occupied by the fluid and
$\partial\Omega$ its boundary.
The last term is zero when $\Omega$ is  closed manifold, or
if we restrict the gauge transformation to those constant
on the boundary. The second term,
\be
{\mathcal W}(g)=\frac{1}{12\pi}\int_\Omega\tr
\left[(g^{-1}dg)^3\right],
\ee
has no reason to vanish, however. It is  the pull-back to
$\Omega$ of an element of,
$H^3_{DR}(G,{\bf Z})$, the third de-Rham cohomology group of
$G$. It can be non-zero
whenever
the gauge transformation  $g(x)$ maps $\Omega$ into a homologically non-trivial
$3$-manifold in
$G$. In particular,  when  $\Omega$ is $S^3$ and the group  $G$ is $SU(2)$, 
then
\be
\frac{1}{12\pi}\int_\Omega\tr
\left[(g^{-1}dg)^3\right]= 2\pi n,
\ee
where  $n$ is the degree  of the map from $S^3\to
SU(2)\simeq S^3$. In this case $k$ has to be  an integer so
that the gauge ambiguity in $C({\mathcal A})$ is $2\pi kn$ and 
$\exp (ikC({\mathcal A}))$ is well-defined.

The Chern-Simons action found by Volovik and Yakovenko has
a coefficient corresponding to $k=1/2$, and so violates the
quantization condition on $k$. It cannot be gauge invariant
on its own. The complete effective action  will be  gauge
invariant, of course, but we must include additional
degrees of freedom to make this  manifest. One of these is
the direction of the vector $\bf d$, which we must
therefore allow to vary.  We parameterize
$\bf d$ in terms of  a group element  $V\in SU(2)$ by setting
\be
 ({\bf d}\cdot\sigma)=  V\sigma_2V^{-1}.
\ee
Then,  
\be
S_{\rm eff}^{\rm spin}({\bf d}, {\mathcal A}) = 
S_{\rm eff}^{\rm spin}(\hat \y, {\mathcal
A}^V)=\frac 12 C({\mathcal A}^V).
\ee 
This expression is clearly invariant under the simultaneous replacement
${\mathcal A}\to {\mathcal A}^U$ and $V\to U^{-1}V$. This  is good, but not
perfect. The problem is that  $V$ is not unique. The vector  $\d$ is
more correctly paramaterized by  elements of the coset
$SU(2)/U(1)$, since we can replace $V$ by
$Ve^{i\sigma_2 \phi}$ without changing $\bf d$. This 
substitution does affect   $C({\mathcal A}^V)$, however. 
Compensating for the effects  of $e^{i\sigma_2 \phi}$ 
requires yet another degree of freedom. We will 
write this as $W=e^{i\sigma_2\chi}$. A  completely gauge invariant action is
then
\be
S_{\rm eff}^{\rm spin}({\bf d}, {\mathcal A}, \chi)= \frac 12 C({\mathcal
A}^{VW}).
\ee
This is manifestly  invariant under the simultaneous
replacement  
\bea
{\mathcal A}\to {\mathcal A}^U, \nonumber\\
V\to U^{-1}Ve^{i\sigma_2 \phi},\nonumber\\
\chi \to \chi-\phi.
\eea

What is the physical  interpretation of  this extra field $\chi$?
This question is easiest to answer if we set $V=I$ on the
boundary, so that $\d=\hat \y$ there. Then,
using the Polyakov-Wiegmann identity (\ref{eq:polyakov}) we can write
\bea
S_{\rm eff}^{\rm spin}({\bf d}, {\mathcal A}, \chi)&=& \frac 12 C({\mathcal
A}^{VW})=
\frac 12 C(A) -\frac{1}{24\pi} \int_\Omega\tr
\left[(V^{-1}dV)^3\right]\nonumber\\
&&\qquad  -\frac{1}{8\pi}\int_{\partial
\Omega}\tr\left\{d W W^{-1}{\mathcal A}\right\}.
\eea 
The second term 
\be
\frac 12 {\mathcal W}(V)= \frac{1}{24\pi} \int_\Omega\tr
\left[(V^{-1}dV)^3\right],
\ee
is precisely the Hopf index   found
by Volovik
and Yakovenko. On a closed manifold, or when ${\bf d}$
is fixed at the boundary, it is equal to $n\pi$ where $n$ labels the
homotopy class of the map ${\bf d}:S^3\to S^2$.  The Berry phase
provided by this term 
makes    a skyrmion soliton in the ${\bf d}$ field into a fermion.  The
third term,
\be
-\frac{1}{8\pi}\int_{\partial
\Omega}\tr\left\{d W W^{-1}{\mathcal A}\right\},
\ee
is equal to  
\be
 -\frac{i}{8\pi}\int_{\partial
\Omega} \tr \left\{d\chi \sigma_2 {\mathcal A}\right\}
=\frac{1}{4\pi}\int dxdt\left\{{\mathcal A}_0^2 \partial_x
\chi - {\mathcal A}_x^2 \partial_t
\chi\right\},
\ee
and  
represents the coupling of the $\chi$ field current to the  $\sigma_2$ 
component of  ${\mathcal A}$. This interaction takes place only on
the boundary, which we have taken to
be the $x$ axis as in section \ref{SEC:planar}.  
This suggests that  the $\chi$ 
field is the bosonized  form  of the complex Weyl fermion
$\Psi_c$ that we
constructed  out of the two Majorana-Weyl edge modes in
section \ref{SEC:planar}. As we noted there, $\Psi_c$
naturally  couples to the  $\sigma_2$  component of ${\mathcal
A}$. 

Gauge invariance tells us that the edge-modes must exist,
but  it
does not determine their dynamics. We 
can, however, add a manifestly gauge invariant boundary term that ensures that  the  $\chi$
field propagates unidirectionallly at speed $c=-\Delta/k_f$. This
term is \cite{stone91}
\be
-\frac c{8\pi} \int dxdt\,\tr\left\{[W^{-1}(\partial_x+{\mathcal
A}_x)W][ W^{-1}(\partial_x+c^{-1}\partial_t+{\mathcal
A}_x+c^{-1}{\mathcal
A}_t)W]\right\}. 
\ee
Including it, and writing $W=e^{i\sigma_2\chi}$, 
the effective action becomes 
\bea
S_{\rm eff}^{\rm spin}({\bf d}, {\mathcal A}, \chi)
&=&\frac 12 C({\mathcal A}) -\frac 12 {\mathcal W}(V) +\frac c{4\pi}\int
dxdt\, \left\{\partial_x
\chi[(c^{-1} \partial_t+\partial_x)\chi]\right\}\nonumber\\
&& + \frac 1{2\pi}\int
dxdt \left\{(c{\mathcal A}^2_x+{\mathcal
A}^2_t)\partial_x\chi\right\} - \frac
1{8\pi}\int dxdt\, \tr \left\{(c{\mathcal A}_x+{\mathcal
A}_t){\mathcal A}_x\right\}.
\nonumber\\
\eea
The  terms containing $\chi$  now constitute the action for a 
chiral boson  interacting with the appropriate chiral component of
the gauge field. The entire expression is invariant under gauge
transformations $U$ which reduce to the form $e^{i\sigma_2\phi}$
on the boundary, and so maintain $\d=\hat \y$ there.
We  note  that  the factor of $1/2$ in the
``level'' $k$  is compensated for by a  factor
$2$ coming from $\tr(\sigma_2^2)=2$, so as  to give the correct 
scale for a chiral boson representing a Weyl fermion.  

Since the action of the $U(1)$ and $SU(2)$ gauge groups
commute with one another, the complete 
gauge invariant effective action containing 
all the low-energy degrees of freedom is the sum
\be
S^{\rm tot}_{\rm eff}(A,\Phi, {\bf d}, {\mathcal A}, \chi)=
2S^{\rm num}_{\rm eff}(A,\Phi) 
+S^{\rm spin}_{\rm eff}({\bf d}, {\mathcal A}, \chi).
\ee
Here the  ``2'' in front  of $S^{\rm num}_{\rm eff}$
comes  from
the two spin components.

\section{Conclusions}
 
We have investigated the gauge invariance of the low-energy
effective action for a $2+1$ dimensional  chiral  superfluid
coupled to external gauge fields. When the order parameter
completely breaks the gauge group, as in the case of  
Abelian particle-number symmetry, the effective action
becomes manifestly gauge invariant as soon as  we include
the bulk Goldstone modes among the dynamical fields. When
the gauge symmetry is not completely broken, as in the case
of  spin-rotation symmetry, we found that additional,
non-Goldstone, degrees of freedom were required for
manifest gauge invariance. These were identified as being
the spin-up and spin-down  Majorona-Weyl edge fermions,
which could be combined to produce a current that  soaks up
the remaining gauge dependence.  

The  consequences  of the effective actions for spin  and
particle-number currents   also differ. The former has a
true spin-Hall effect, with a dissipationless  spin current
proportional  the spin ``electric'' field. The latter, we
have argued, has only a ``mock'' Hall effect, the induced
current being proportional to the change in density, and
not to the external field causing the change.          

\section{Acknowledgments}

This work was supported by the National Science Foundation
under grant NSF-DMR-01-32990. MS would like to thank Eduardo
Fradkin, Rinat Kedem and Eddy Ardonne for useful
conversations, and Grisha Volovik for a valuable e-mail
exchange.

\appendix

\section{Dirac equation} 

The twisted-mass Dirac equation that results from our
two-dimensional problem is a standard illustration of the theory
of fractional charge \cite{goldstone-wilczek}. 
We review it here so as to make clear the contribution of
the extended scattering states to the boundary current. 

We consider the one-dimensional Dirac Hamiltonian
\be
\hH= -i\tau_3\partial_x +\Delta\tau_1e^{-i\tau_3\phi(x)}
=\left[\matrix{-i\partial_x & \Delta e^{i\phi(x)}\cr
                \Delta e^{-i\phi(x)}& i\partial_x }\right],
\ee
where $\Delta$ is a constant.

We will compute the extra particle number that is
accumulated in the vicinity of $x=0$ when 
$\phi$ is discontinuous,
jumping  abruptly from  
$\phi=\phi_L$ when   $x<0$ to $\phi=\phi_R$ when $x>0$. 
Suppose  the eigenstates of $\hH$ are $\chi_n$ with energy
$E_n$. 
The ground-state number density is  
\be
\vev{\psid \psi(x)}= \sum
|\chi_n(x)|^2,
\ee
where the sum is over  occupied states, {\it i.e.\/} those
with $E_n<0$. Because the
sum of  $|\chi_n|^2$ over {\it all\/} states is independent
of $\phi$ by completeness, we can equally well write
\be
\vev{\psid \psi(x)}= \hbox{const.}- \sum_{E_n>0}
|\chi_n(x)|^2,
\ee
and this form is slightly more convenient.
We will show that 
\bea
Q= \int \vev{\psid \psi(x)}dx  
&=& \frac 1{2\pi} (\phi_R-\phi_L),\quad  
0<(\phi_R-\phi_L)<\pi,\nonumber\\
&=&  \frac 1{2\pi} (\phi_R-\phi_L)-1,\quad 
\pi<(\phi_R-\phi_L)<2\pi.
\eea
 
When $\phi$ is constant we have a continuum of positive and negative
energy 
eigenfunctions
\be    		
\psi^\phi_{k,E} =  e^{i\sigma_3
\phi/2}\frac{1}{2\sqrt{E(E+\Delta)}} 
\left[\matrix{E+k+\Delta\cr E-k+\Delta}\right]e^{ikx},
\ee
where $E(k)=\pm \sqrt{k^2+\Delta^2}$. 
With the discontinuity  present, we will have scattering solutions
\bea
\psi &=& a_L^{(in)}\psi^{\phi_L}_{k,E}+
a_L^{(out)}\psi^{\phi_L}_{-k,E},\quad x<0,\nonumber\\
&=& a_R^{(in)}\psi^{\phi_R}_{-k,E}+
a_R^{(out)}\psi^{\phi_R}_{k,E},\quad x>0.\nonumber
\eea
The function $\psi$  must be continuous at $x=0$, and from
this condition we obtain the  $S$-matrix relation
\be
\left[\matrix{a_L^{(out)}\cr a_R^{(out)}}\right]=
\left[\matrix{t&r\cr r& t}\right]
\left[\matrix{a_R^{(in)}\cr a_L^{(in)}}\right], 
\ee
where
\bea
t(k,E) &=& \frac{1}{\cos(\Phi/2) - (iE/k) 
\sin(\Phi/2)},\nonumber\\
r(k,E) &=& \frac{i(\Delta/k)\sin(\Phi/2)}
{\cos(\Phi/2) - (iE/k) 
\sin(\Phi/2)}.
\eea
Here $\Phi$ is shorthand for $\phi_L-\phi_R$.

In addition to the continuum states there is  also a single bound state
\bea
\psi_0 &\propto & e^{i\sigma_3 \phi_R/2} 
\left[\matrix{E_0+i\kappa+\Delta\cr
E_0-i\kappa+\Delta}\right]e^{-\kappa x},\quad x>0,\nonumber\\
&\propto &  e^{i\sigma_3 \phi_L/2} 
\left[\matrix{E_0-i\kappa+\Delta\cr
E_0+i\kappa+\Delta}\right]e^{\kappa x}.\quad x<0.\nonumber
\eea
The bound-state energy $E_0$ is also determined by 
continuity at $x=0$, which  requires 
\bea
E_0&=& \Delta \cos(\Phi/2),\nonumber\\
\kappa &=& \Delta \sin(\Phi/2).\nonumber
\eea
These formula are valid for  $0<\Phi<2\pi$, where $\kappa$
is positive, and  extend outside that interval with period $2\pi$. 
At $\Phi=0$ there is no bound state. As
$\Phi$ increases, a bound state peels off the upper
continuum. It passes through $E=0$ at $\Phi=\pi$, and
merges with the lower continuum as $\Phi$ reaches $2\pi$.
If $\Phi$ increases beyond $2\pi$, the process repeats
with another state peeling off the upper continuum.
Thus each $2\pi$ twist in $\Phi$ results in the net
transfer of one state from the upper continuum to the lower.

Using the relation between $E_0$ and $\kappa$, we 
can write the normalized bound
state as 
\be
 \chi_0=
 \sqrt{\frac{\kappa}{2}}\left[\matrix{e^{i(\phi_L+\phi_R)/4}\cr
e^{-i(\phi_L+\phi_R)/4}}\right]e^{-\kappa|x|}.
\ee
A complete set of states comprises $\chi_0$ together with 
 \be
 \chi_{k,E} = \left\{\matrix{\psi^{\phi_L}_{k,E}+r(k,E)
 \psi^{\phi_L}_{-k,E},&\quad x<0,\cr
  t(k,E) \psi^{\phi_R}_{k,E},&\quad x>0;}\right.\quad  k>0.
  \ee
  \be
 \chi_{k,E} = \left\{\matrix{t(k,E)\psi^{\phi_L}_{k,E},&\quad
 x<0,\cr
 \psi^{\phi_R}_{k,E} + r(k,E)\psi^{\phi_R}_{-k,E},&\quad
 x>0}\right.\quad  k<0
 \ee
 These basis states therefore switch from a wave incident from the
 left to one incident from the right as $k$ changes sign.

\begin{figure}
\includegraphics[width=3.0in]{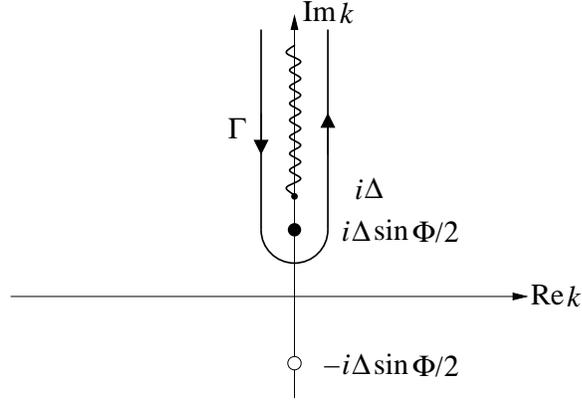}
\caption{\label{cut} The contour $\Gamma$,
showing the cut starting at $ik=\Delta$. When $0<\Phi<\pi$
there is  pole at
$k=i\Delta\sin\Phi/2$, and a pole on the second sheet at
$k=-i\Delta\sin\Phi/2$. }
\end{figure}

After using $|r|^2+|t|^2=1$, and the explicit
form of the free eigenfunctions, we find  
\be
\sum_{E>\Delta}
|\chi_E(x)|^2
=\int_{-\infty}^{\infty}\frac {dk}{2\pi}
|\chi_{E,k}(x)|^2
= \hbox{const.}+\int_{-\infty}^{\infty}\frac {dk}{2\pi}r(k,E)
\left(\frac
{\Delta}{k}\right) e^{2ik|x|}.
\label{EQ:upper_continuum}
\ee
Here the ``const.'' refers to terms that are independent of
$\phi_{R,L}$. 
We can  improve the numerical convergence of the
integral  by using Jordan's lemma to push the contour of
integration $\Gamma$ into the upper half plane, as shown in
figure \ref{cut}. 

%\vbox{
%\vskip 20 pt
%\centerline{\epsfxsize=3.0in\epsffile{cut.eps}}
%{\narrower\smallskip\noindent{\sl The contour $\Gamma$,
%showing the cut starting at $ik=\Delta$. When $0<\Phi<\pi$
%there is  pole at
%$k=i\Delta\sin\Phi/2$, and a pole on the second sheet at
%$k=-i\Delta\sin\Phi/2$. }
%}}

We observe  that the   reflection
coefficient has
a cut
running from $k=i\Delta$ to $i\infty$, and that  for
$0<\Phi<\pi$ there is  a pole in the upper half plane 
at $k=i\Delta \sin\Phi/2$. At $\Phi=\pi$ the pole merges
with the cut. For $\pi<\Phi<2\pi$ the pole is apparently
returning towards the real axis again, but a more careful
investigation shows that it is now 
on the second sheet, and no longer contributes. (At the same time a
pole at $-i\Delta \sin\Phi/2$ has emerged onto the first
sheet in the lower half plane. This is below the real axis,
however, and also does not contribute.) 

The integral in ({\ref{EQ:upper_continuum}) then becomes   
\be
-\Delta \sin\Phi/2 e^{-2\Delta
|\sin\Phi/2||x|} + \int_\Delta^\infty \frac{dk}{2\pi}
\frac{\Delta^2\sin\Phi}{\kappa^2-\Delta^2\sin^2(\Phi/2)}\frac
1{\sqrt{\kappa^2-\Delta^2}} \kappa e^{-2\kappa|x|}.
\label{EQ:charge_distribution}
\ee
The first term,  the  pole contribution,
is only present if  $0<\Phi< \pi$.

There does not seem to be a closed-form expression for the
integral in (\ref{EQ:charge_distribution}), but if we first
integrate  over $x$ to
get the total charge, we end up with an elementary integral 
\be
\int_\Delta^\infty\frac{dk}{2\pi}
\frac{\Delta^2\sin\Phi}{\kappa^2-\Delta^2\sin^2(\Phi/2)}\frac
1{\sqrt{\kappa^2-\Delta^2}}= \frac {\Phi}{2\pi},\quad
-\pi<\Phi<\pi.
\ee
This expression extends to a $2\pi$ periodic function of
$\Phi$, and  the integral is therefore discontinuous at odd multiples
of $\pi$. After we include the pole contribution, which is
discontinuous at {\it all\/} multiples of $\pi$, we find 
that the total continuum contribution  is discontinuous only at
$\Phi=0$ (mod $2\pi$). Thus  
 \be
\int_{-\infty}^{\infty}\frac {dk}{2\pi}|\chi_{E,k}(x)|^2
= \hbox{const.}+ \frac {\Phi}{2\pi},\quad 0<\Phi<2\pi.
\label{EQ:upper_charge}
\ee
The discontinuity at $\Phi=0$ (mod $2\pi$) is due to the
sudden loss of the bound state from the upper continuum. As the bound
state makes its way to the lower continuum, the spectral
weight in the upper continuum is gradually recovered.  

We can now apply these results to compute
\be
Q= \int   \sum_{E_n<0}|\psi_n(x)|^2dx
\ee
We observe that our sum over the positive continuum is equal
to minus the sum over the negative continuum together with  the bound
state. The bound state has negative energy, however,  if
$\pi<\Phi<2\pi$ (mod $2\pi$). If we are computing the
ground state charge, and  if
$0<\Phi<\pi$  (mod $2\pi$), we should reduce the sum by
unity. We also note that the constant in
(\ref{EQ:upper_charge}) is fixed by the requirement that the
the accumulated charge be zero when $\phi_L=\phi_R$. Thus 
\bea
Q &=& \frac 1{2\pi} (\phi_R-\phi_L),\quad  
0<(\phi_R-\phi_L)<\pi,\nonumber\\
&=&  \frac 1{2\pi} (\phi_R-\phi_L)-1,\quad
\pi<(\phi_R-\phi_L)<2\pi.
\label{EQ:lost_particle}
\eea
This result  repeats with $2\pi$ periodicity.

This expression for $Q$  is consistent with the basic result
of Goldstone and Wilczek \cite{goldstone-wilczek}, that a total charge of 
\be
\int \frac 1{2\pi} \partial_x \phi\, dx= \frac
1{2\pi}(\phi_R-\phi_L)
\label{EQ:goldstone}
\ee
is drawn into a region as  the phase $\phi$ is slowly twisted. 
The difference between (\ref{EQ:lost_particle}) and
(\ref{EQ:goldstone}) arises because the latter  does not keep track of
individual particles that are lost to the reservoir when
their energy exceeds the chemical potential. 

In a superfluid the role of charge and current is
interchanged. The fractional charge, multiplied by $k_f$ and divided
by two,  gives the momentum density, or equivalently, the
mass current \cite{stone-gaitan}.

%\section{References}
%comment out for revtex

\end{document}